# Unbiased crystal structure prediction of NiSi under high pressure

Pavel N. Gavryushkin,[a b *] Zakhar I. Popov,[c] Konstantin D. Litasov,[a b], Alex N. Gavryushkin[d]

[a] V.S. Sobolev Institute of Geology and Mineralogy, Russian Academy of Science, Siberian Branch, Russia,
[b] Novosibirsk State University, Novosibirsk, Russia,
[c] Kirensky Institute of Physics, Russian Academy of Science, Siberian Branch, Russia,
[d] Department of Computer Science University of Auckland, New Zealand

Correspondence e-mail: p.gavryushkin@gmail.com

**Abstract**

Based on the unbiased structure prediction, we showed that the stable form of NiSi compound under the pressure of 100 and 200 GPa is the *Pmmn*-structure. Furthermore, we discovered a new stable phase—the deformed tetragonal CsCl-type structure with $a$ = 2.174 Å and $c$ = 2.69 Å at 400 GPa. Specifically, the sequence of high-pressure phase transitions is the following: the *Pmmn*-structure – below 213 GPa, the tetragonal CsCl-type – in the range 213-522 GPa, and cubic CsCl – higher than 522 GPa. As the CsCl-type structure is considered as the model structure of FeSi compound at the conditions of the Earth's core, this result implies restrictions on the Fe-Ni isomorphic miscibility in FeSi.

**Introduction**

Nickel silicide is isostructural to silicides of other transition metals (MnSi, FeSi, CoSi) (Toman, 1951), crystallizes in the MnP-structure under ambient conditions, and has a wide range of practical applications as the material for contacts in microelectronics (Lavoie *et al.*, 2006). Although high pressure polymorphs of NiSi, which is the object of study in this paper, have currently not found practical applications, the information about its structures can be useful for the analysis of metastable states of NiSi. These states are formed, for example, during the crystallization of NiSi thin films on the silicon substrate (d'Heurle *et al.*, 1984, De Keyser *et al.*, 2008).

In the same way as other silicides of transition metals, NiSi goes through a series of phase transitions under pressure which finishes in the CsCl-type structure (Lord *et al.*, 2012, Zhang & Oganov, 2010). The electronic structures of CsCl-phases of FeSi and CoSi attract a special attention due to the fact that these compounds belong to the semiconductors with narrow band gap (Al-Sharif *et al.*, 2001, Zhao *et al.*, 2011, Acun & Soyalp, 2012, Jin *et al.*, 2013).

Among other silicides of transition metals, the silicides of nickel and iron are of interest for the Earth sciences because both iron and nickel are the most important compounds of the Earth's core and silicon is one of the candidates for being a light element in the core (Poirier, 1994). Despite of these arguments, the presence of nickel silicide in the form of separate phase in the Earth's core seems to be unlikely. The information about its high pressure phases is important for assessing the Fe-Ni isomorphism in the FeSi structure. In this paper, we will discuss the possibility of such an isomorphism based on the information about new high pressure phases of NiSi.

We are motivated to perform the calculations on structure prediction of this compound by the complexity of NiSi phase diagram and the difficulties in designing and interpreting the experiments with the compound. Our calculations were fully *ab-initio* and unbiased. Structure predictions based on known AB structures have been done in details before (Wood *et al.*, 2013, Vocadlo *et al.*, 2012, Lord *et al.*, 2012). We also would like to mention that in the last decade *ab-initio* structure prediction methods have become a full-value instrument of crystallographic investigations and crystal structure determinations (Oganov, 2011), especially under conditions when obtaining the high quality experimental data is difficult or impossible (Oganov & Glass, 2006, Oganov & Ono, 2004). This is applicable in full to the experiments with NiSi at high pressures.

As our interest to NiSi compound was mainly geological, we restricted the pressure range to 100-400 GPa with the main focus being on pressures of 300-400 GPa, i.e. the pressures of the Earth's core. The purpose of our investigations can hence be formulated as an unbiased search for structures with the energy lower than that of known CsCl-type structure of NiSi under the pressures of the Earth's core.

**Calculation details**

Crystal structure prediction has been carried out using the evolutionary algorithms implemented in USPEX package (Oganov & Glass, 2006, Lyakhov *et al.*, 2010, Lyakhov *et al.*, 2013, Glass *et al.*, 2006, Oganov *et al.*, 2006). No restrictions on the symmetry or other structure parameters have been imposed. The search was performed at 100, 200, 300, 400 GPa for one, two, three, and four formula units for each pressure. The temperature was 0 K in all calculations. During the search, structures relaxed with VASP 5.3 code (Vienna ab initio simulation package) (Kresse & Joubert, 1999, Kresse & Furthmuller, 1996) in the framework of density functional theory (DFT) (Kohn & Sham, 1965) using the plane wave basis set and the projector augmented wave (PAW) method (Blöchl, 1994, Kresse & Joubert, 1999). Exchange-correlation effects were taken into account in the generalized gradient approximation (GGA) by the Perdew–Burke–Ernzerhof (PBE) functional (Perdew *et al.*, 1996). The following orbitals were treated as valence states: 3s and 3p for Si; 3p, 3d and 4s for Ni. An energy cut off of 480 eV was used, which is nearly 30% higher than maximum energy cut off from pseudopotentials. During relaxation, atoms were shifted until forces acting on them became less than $0.01 eV Å^{-1}$. The first Brillouin zone was sampled [13*13*13] using the Monkhorst–Pack scheme (Monkhorst & Pack, 1976).

Dynamical stability calculations were carried out with Phonopy package (Togo *et al.*, 2008).

To test the used methodology, the nontrivial MnP-type structure (Toman, 1951) was predicted for NiSi at 0 GPa. The test on estimating the pressure error shows that the error is less than 3 GPa.

**Results and discussion**

The calculations at 100 and 200 GPa reveal the *Pmmn* structure as one with the lowest enthalpy. This result is in a full agreement with the experimental and theoretical work of Wood and co-authors (Wood *et al.*, 2013). The unit cell parameters and atomic coordinates predicted in our calculations are identical to those obtained in that work. The symmetry analysis of *Pmmn*-structure shows that increasing the tolerance on symmetry leads to the *P4/nmm* space group (Vocadlo *et al.*, 2012). However, the enthalpy of *Pmmn* structure is sufficiently lower than that of its *P4/nmm* arisotype: the enthalpy difference at 200 GPa is almost equal to 0.3 eV. This is a good example of how an improper symmetry analysis can lead to the most favorable structure being missed. *Our results in the pressure range of 100-200 GPa tell that despite the ideas about the most favorable structure at these pressures being changed several times (Wood et al., 2013), finally founded Pmmn structure is actually the most favorable one at 0K.*

We discovered a new structure at the pressure ranges of 300 and 400 GPa. The new structure is the tetragonal distortion of CsCl-type structure, see Fig.1. The calculation of the phonon dispersion shows that this new phase is dynamically stable (the phonon dispersion curves are presented in the *supplementary materials*). The ideal CsCl-type structure was also found during the calculations. The predicted unit cell parameters of this structure are in good agreement with the experimental ones: P=101 GPa, *a*(predicted)=2.573 Å, *a*(experimental)=2.579 Å (Lord *et al.*, 2012). The enthalpy of a tetragonal CsCl-type structure is lower than that of the cubic one in the pressure range of 0-522 GPa. The maximum difference of enthalpies is 0.05 eV at 130 GPa (Fig.1). The *a/c* ratio for the tetragonal phase is approximately 0.8 and remains constant throughout the pressure range. The unit cell parameters at 400 GPa were calculated as *a*=2.174 Å and *c*=2.69 Å. For comparison, the *a*-axis of the cubic phase at the same pressure is 2.332 Å. The cubic phase is characterized by slightly higher compressibility than the tetragonal phase: at 0 GPa the volume of the tetragonal phase is less by 0.7% than the volume of the cubic phase, at 130 GPa their volumes are equal, and at 400 GPa the volume of the cubic phase is less by 0.2% than that of the tetragonal phase. The dependence of unit cell parameters and enthalpy on pressure is plotted in the *supplementary materials*.

*The geological meaning of the tetragonal CsCl-type structure is that the presence of this phase at the conditions of the Earth's core can imply restrictions on isomorphic miscibility of (Fe,Ni) in the CsCl-type polymorph of FeSi (Belonoshko et al., 2009), because the end members are no more isostructural as it was assumed earlier.*

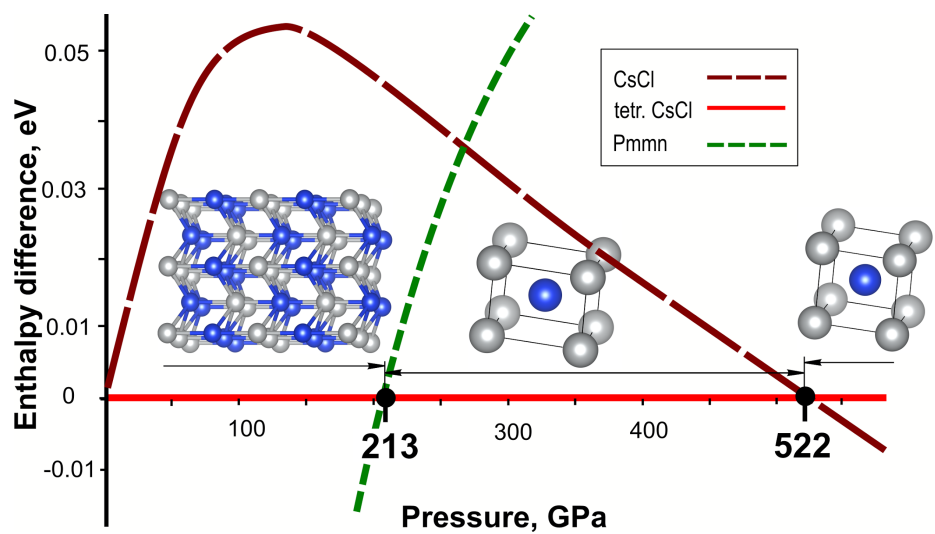

Fig.1. Enthalpies of high pressure NiSi-phases normalized by enthalpy of tetragonal CsCl phase.

*Based on our results on ab-initio crystal structure prediction, the following sequence of high-pressure phase transitions can be proposed: the Pmmn structure should be stable at pressures lower than 213 GPa, the tetragonal CsCl – at pressures from the range of 213-522 GPa, and cubic CsCl—at pressures above 522 GPa.*

A distinct crystallogrphic feature of all four high-pressure phases of NiSi (ε-FeSi, *Pmmn*, tetragonal CsCl, and ideal CsCl) is that crystallographic orbits corresponding to both types of atoms are Euclidean-equivalents, i.e. both types of atoms are characterized by the same coordination polyhedrons packed in the same manner. However, this is not true for the ambient MnP-phase of NiSi, in which Ni and Si atoms form substantially different nets. This difference can be an indication of fundamental changes in the character of the Ni–Si chemical bond which occurs under compression.

**Comparison with experimental data**

The CsCl-type phase of NiSi was synthesized by (Lord *et al.*, 2012) in laser-heated diamond anvill cell (DAC) experiments at pressure of 49 GPa and temperatures of 1900 K. To explain why the tetragonal phase was not synthesized in these experiments, we propose two major reasons. The first reason is related to the stabilization effect of temperature. The contribution of the entropy term to the Gibbs energy can be substantial under such a high temperature and this can change the relations between the phases determined at 0 K (Belonoshko *et al.*, 2003). As the difference in enthalpy between the cubic and tetragonal phases is relatively small (less than 0.05 ev), this explanation is natural. The second reason is related to experimental difficulties. The CsCl-type phase was identified in the diffraction patterns along with four other phases. Some phases, which were not recognized in those DAC experiments, were later synthesized in a large volume multianvil apparatus (Wood *et al.*, 2013). The experimentally determined unit cell parameters of the cubic phase are slightly deviated from the ideal ratio of the cubic symmetry towards the tetragonal symmetry. Hence, we suggest that the synthesis of the tetragonal phase is possible in future high-pressure experiments with or without heating.

**Acknowledgements**

We thank Information Technology Centre of Novosibirsk State University for providing access to the cluster computational resources. The research was financially supported by the Russian Science Foundation (No 14-17-00601) and performed under the program of the Ministry of Education and Science of Russia (project № 14.B25.31.0032). The work of ZP is supported by the Leading Science School programme (No NS-2886.2014.2).

**References**

Acun, A. D. & Soyalp, F. (2012). *Philosophical Magazine* **92**, 635-646.
Al-Sharif, A., Abu-Jafar, M. & Qteish, A. (2001). *Journal of Physics: Condensed Matter* **13**, 2807.
Belonoshko, A. B., Ahuja, R. & Johansson, B. (2003). *Nature* **424**, 1032-1034.
Blöchl, P. E. (1994). *Phys Rev B* **50**, 17953.
d'Heurle, F., Petersson, C., Baglin, J., La Placa, S. & Wong, C. (1984). *J Appl Phys* **55**, 4208-4218.
De Keyser, K., Van Bockstael, C., Detavernier, C., Van Meirhaeghe, R., Jordan-Sweet, J. & Lavoie, C. (2008). *Electrochemical and Solid-State Letters* **11**, H266-H268.

Glass, C. W., Oganov, A. R. & Hansen, N. (2006). *Computer Physics Communications* **175**, 713-720.
Jin, Y.-Y., Kuang, X.-Y., Wang, Z.-H. & Huang, X.-F. (2013). *High Pressure Research* **33**, 15-26.
Kohn, W. & Sham, L. J. (1965). *Physical Review* **140**, A1133.
Kresse, G. & Furthmuller, J. (1996). *Comput. Mater. Sci.* **6**, 15-50.
Kresse, G. & Joubert, D. (1999). *Phys Rev B* **59**, 1758-1775.
Lavoie, C., Detavernier, C., Cabral Jr, C., d'Heurle, F., Kellock, A., Jordan-Sweet, J. & Harper, J. (2006). *Microelectronic engineering* **83**, 2042-2054.
Lord, O. T., Vocadlo, L., Wood, I. G., Dobson, D. P., Clark, S. M. & Walter, M. J. (2012). *J. Appl. Crystallogr.* **45**, 726-737.
Lyakhov, A. O., Oganov, A. R., Stokes, H. T. & Zhu, Q. (2013). *Computer Physics Communications* **184**, 1172-1182.
Lyakhov, A. O., Oganov, A. R. & Valle, M. (2010). *Computer Physics Communications* **181**, 1623-1632.
Monkhorst, H. J. & Pack, J. D. (1976). *Phys Rev B* **13**, 5188.
Oganov, A. R. (2011). *Modern methods of crystal structure prediction*. WILEY-VCH.
Oganov, A. R. & Glass, C. W. (2006). *J. Chem. Phys.* **124**.
Oganov, A. R., Glass, C. W. & Ono, S. (2006). *Earth Planet Sc Lett* **241**, 95-103.
Oganov, A. R. & Ono, S. (2004). *Nature* **430**, 445-448.
Perdew, J. P., Burke, K. & Ernzerhof, M. (1996). *Physical review letters* **77**, 3865.
Poirier, J.-P. (1994). *Physics of the earth and planetary interiors* **85**, 319-337.
Togo, A., Oba, F. & Tanaka, I. (2008). *Phys Rev B* **78**, 134106.
Toman, K. (1951). *Acta Crystallographica* **4**, 462-464.
Vocadlo, L., Wood, I. G. & Dobson, D. P. (2012). *J. Appl. Crystallogr.* **45**, 186-196.
Wood, I. G., Ahmed, J., Dobson, D. P. & Vocadlo, L. (2013). *J. Appl. Crystallogr.* **46**, 14-24.
Zhang, F. W. & Oganov, A. R. (2010). *Geophys Res Lett* **37**.
Zhao, K. M., Jiang, G. & Wang, L. L. (2011). *Physica B* **406**, 363-367.